\documentclass[pra,twocolumn,showpacs,preprintnumbers,superscriptaddress]
{revtex4}

\usepackage{times}
\usepackage{bm}
\usepackage{graphicx}
\usepackage{amsbsy}
\usepackage{amsmath}
\usepackage{amsfonts}
\usepackage{amsthm}

\begin{document}

\theoremstyle{plain}
\newtheorem{theorem}{Theorem}
\newtheorem{lemma}[theorem]{Lemma}
\newtheorem{corollary}[theorem]{Corollary}
\newtheorem{proposition}[theorem]{Proposition}
\newtheorem{conjecture}[theorem]{Conjecture}

\theoremstyle{definition}
\newtheorem{definition}[theorem]{Definition}

\theoremstyle{remark}
\newtheorem*{remark}{Remark}
\newtheorem{example}{Example}

\title{Feed-forward control scheme generate Bell states and three-qubit W-type states
when qubits passes through decoherence channel}

\author{Satyabrata Adhikari}
\thanks{tapisatya@gmail.com}
\affiliation{Birla Institute of Technology Mesra, Ranchi-835215,
India}
\date{\today}

\begin{abstract}
It is known that maximally entangled Bell state and three-qubit
W-type states are very useful in various quantum information
processing task. Thus the problem of preparation of these type of
states is very important in quantum information theory. But the
factor which prohibit the generation of the above mentioned pure
states shared between two and three distant partners is
decoherence. When we send one qubit, from a two qubit state,
through decoherence channel like amplitude damping channel, the
purity of the qubit is lost and it ends up with a mixed state.
Therefore it is very difficult to keep the pure maximally
entangled state in a maximally entangled pure state or in an
entangled state with high entanglement. In this work we have
provided a method by which one can generate experimentally a
maximally entangled Bell states shared between distant parties
with a non-zero probability when a qubit, from a two qubit general
state, passes through decoherence channel. Therefore, despite of
the fact that qubit is interacting with the noisy channel, we are
able to generate Bell state shared between two distant partners.
Further, we have shown that it is possible to generate pure
three-qubit W-type states shared between three distant partners
using economical quantum cloning machine and weak-measurement
based feed-forward control scheme, even though the second and
third qubit is interacting with the noisy channel. Lastly, we have
shown that the generated three-qubit W-type states can be used in
teleporting one of the two non-orthogonal states.
\end{abstract}


\maketitle

\section{Introduction}
Quantum entanglement \cite{epr} is such a feature of quantum
mechanics which has no classical analogue. The entangled resource
state may have bipartite or multipartite entanglement. Quantum
information processing tasks which were initially introduced for
bipartite system can be extended to multipartite system later. In
bipartite system either the state is entangled or separable and if
the state is entangled then it is genuine entangled state. Bell
states are maximally entangled states and it is very useful in
various quantum information processing tasks. Unlike two qubit
states, the multipartite entangled states can be classified
according to various schemes \cite{gw,ss}. Three-qubit states have
been classified according to stochastic local operation and
classical communication (SLOCC) into six categories. Two of these
categories have genuine tripartite entanglement, viz. GHZ-states
and W-states \cite{dvc}. 3-tangle is one of the measures by which
one can distinguish GHZ-states and W-states \cite{es}. For
GHZ-states, 3-tangle is non-zero while for W-states, it is zero.
Previously it was known that W-type states cannot be used for
teleportation and superdense coding but Agrawal and Pati \cite{ap}
introduced a new class of W-type states
which are useful in teleportation and superdense coding.\\
Since entangled states play a major role in various quantum
information processing tasks such as quantum teleportation
\cite{bbcjpw}, superdense coding \cite{bw}, remote state
preparation \cite{pati,bvsstw}, secret sharing \cite{hbb},
telecloning \cite{mpv} and quantum cryptography
\cite{bb,ekert,grtz} so its generation and manipulation is very
important in quantum information theory. There are schemes based
on unitary dynamics for the generation of entangled states
\cite{neeley,smsn}. Beside these schemes there are methods of
generating entanglement by measurements \cite{hm,lgb}. It is very
difficult to store the generated entangled states by
measurement-alone approach. This problem can be resolved by the
technique of quantum feedback control \cite{wm,ch,fwdxl,tj}. An
experiment was proposed to generate and stabilize entanglement
between two qubits in circuit QED \cite{sgsm,liu}. An experimental
demonstration of the generation of superconducting two-qubit Bell
state by feedback based on parity measurements is presented in
\cite{riste}. S-Y Huang et.al. \cite{huang} recently presented a
simple measurement and feedback control scheme feasible with
current circuit QED technology to produce and stabilize the W
state
$\frac{1}{\sqrt{3}}(|100\rangle+|010\rangle+|001\rangle)$.\\
In this work our task is to produce two-qubit Bell states shared
between two distant parties and three-qubit W-type states shared
between three parties. We assume here that a general two qubit
entangled state is prepared in the Alice's laboratory. When Alice
want to share one qubit, from the general two qubit state, with
Bob, she has to sent the qubit through decoherence channel. In
general when qubit is interacted with the noisy environment, it
loses its purity and becomes a mixed state. Although one qubit is
sent through decoherence channel, we have shown that Alice and Bob
manage to share a maximally entangled Bell states by following C-Q
Wang et.al. \cite{wang} weak-measurement-based feed-forward
control scheme. Starting from the generated Bell state shared
between two distant partners Alice and Bob, a three-qubit W-type
state is prepared by using economical quantum cloning machine and
weak-measurement-based feed-forward scheme. The generated three
qubit state is not a mixed state but a pure three-qubit W-type
state, which is shared between three distant partners Alice, Bob
and Charlie. Later we have shown that the generated pure
three-qubit W-type state can be used in
the teleportation of two non-orthogonal states.\\
The plan of the paper is as follows: In section 2, we revisit C-Q
Wang et.al. weak-measurement-based feed-forward control scheme. In
this scheme, authors showed that how one can protect the purity of
the qubit when passing through decoherence channel. In section 3,
we have used the concept discussed in section 2 to generate
two-qubit Bell states shared between sender and the receiver
located far away from each other, when a qubit passes through
amplitude damping channel. In section 4, we have shown that a
particular form of three-qubit W-type state is produced by
economical quantum cloning machine and weak-measurement-based
feed-forward control scheme. In section 5, the generated
three-qubit W-type state is used to teleport two non-orthogonal
states. Finally, we conclude in section 6.
\section{C-Q Wang et.al. weak-measurement-based feed-forward
control scheme} C-Q Wang et.al. \cite{wang} introduced a
feed-forward control scheme to protect an unknown quantum state.
The scheme is based on one complete pre-weak measurement, two
incomplete post-weak measurements and two feed-forward operations
and their reversals. The scheme of protecting unknown quantum
states when it passes through decoherence channel, goes as
follows: Let us consider an unknown quantum state
\begin{eqnarray}
|\psi\rangle=\alpha|0\rangle +\beta
|1\rangle,~~~~~|\alpha|^{2}+|\beta|^{2}=1
\label{unknownstate}
\end{eqnarray}
Firstly, we perform a pre-weak measurement on the state
$|\psi\rangle$ in (\ref{unknownstate}) before it passes through
the noisy channel. Pre-weak measurement can be chosen as
$\Pi_{1}=M_{1}^{\dagger}M_{1}$ and $\Pi_{2}=M_{2}^{\dagger}M_{2}$,
and $M_{1}$ and $M_{2}$ are given by
\begin{eqnarray}
M_{1}=\begin{pmatrix}
  \sqrt{p} & 0 \\
  0 & \sqrt{1-p}
\end{pmatrix}, M_{2}=\begin{pmatrix}
  \sqrt{1-p} & 0 \\
  0 & \sqrt{p}
\end{pmatrix} \label{pre-weakmeasurement}
\end{eqnarray}
where $p$ is the pre-weak measurement strength and
$\sum_{i=1}^{2}M_{i}^{\dagger}M_{i}=I$.\\
If the measurement outcome is $M_{1}$ then we adopt the
feed-forward operation $F_{1}=I$ but if the measurement outcome is
$M_{2}$, we adopt the feed-forward operation $F_{2}=\sigma_{x}$.
In this work we explain the whole protocol when the measurement
outcome is $M_{1}$. In a similar fashion, we could demonstrate the
whole protocol with measurement outcome $M_{2}$ also.\\
The occurrence of the measurement outcome $M_{1}$ reduces the
state $|\psi\rangle$ to
\begin{eqnarray}
|\psi\rangle_{M_{1}}=\frac{M_{1}|\psi\rangle}{\sqrt{\langle\psi|\Pi_{1}|\psi\rangle}}=\frac{1}{N_{M_{1}}}(\alpha\sqrt{p}|0\rangle
+\beta\sqrt{1-p} |1\rangle) \label{m1state}
\end{eqnarray}
where
$N_{M_{1}}=\langle\psi|\Pi_{1}|\psi\rangle=|\alpha|^{2}p+|\beta|^{2}(1-p)$.
We perform the feed-forward operation $F_{1}=I$ on
$|\psi\rangle_{M_{1}}$ which keeps the state as it is. Qubit
(\ref{m1state}) is then passed through the noisy channel for a
period of $\tau$. The qubit is then no longer pure because of
energy relaxation with the rate $\Gamma$. To keep the qubit in a
pure state, we disentangle the relaxation into "jump" and "no
jump" scenarios. When the qubit is passing through the amplitude
damping channel, the qubit trajectories is divided into two parts:
(i) the qubit is jumping into the state $|0\rangle$ with the
"jump" probability
$P^{j}=N_{M_{1}}|\beta|^{2}(1-e^{-\Gamma\tau})$. (ii) "no jumping"
state of the qubit is
\begin{eqnarray}
|\psi\rangle_{nj}=\frac{1}{\sqrt{P^{nj}}}(\alpha\sqrt{p}|0\rangle
+\beta\sqrt{1-p}e^{\frac{-\Gamma\tau}{2}} |1\rangle)
\label{nojumpingstate}
\end{eqnarray}
where $P^{nj}=|\alpha|^{2}p+|\beta|^{2}(1-p)e^{-\Gamma\tau}$. Then
the reversed feed-forward operation $F_{1}=I$ retain the state
$|0\rangle$ and $|\psi\rangle_{nj}$. Lastly, we measure the qubit
by post-weak measurement $\bigwedge=O_{1}^{\dagger}O_{1}$, where
\begin{eqnarray}
O_{1}=\begin{pmatrix}
  \sqrt{1-p_{1}} & 0 \\
  0 & 1
\end{pmatrix}, \label{post-weakmeasurement}
\end{eqnarray}
$p_{1}$ is the post-weak measurement strength.\\
The measured state from the "jumping" trajectory is $|0\rangle$
with probability
$P^{jN_{1}}=N_{M_{1}}|\beta|^{2}(1-e^{-\Gamma\tau})(1-p_{1})$. The
measured state from the "no jumping" trajectory is given by
\begin{eqnarray}
|\psi\rangle_{njn1}=\frac{1}{\sqrt{P^{njN_{1}}}}(\alpha\sqrt{p}\sqrt{1-p_{1}}|0\rangle
+\beta\sqrt{1-p}e^{\frac{-\Gamma\tau}{2}} |1\rangle)
\label{nojumpingstaten1}
\end{eqnarray}
where
$P^{njN_{1}}=|\alpha|^{2}p(1-p_{1})+|\beta|^{2}(1-p)e^{-\Gamma\tau}$.\\
If we choose the post-weak measurement strength as
\begin{eqnarray}
p_{1}=1-\frac{(1-p)e^{-\Gamma\tau}}{p}
 \label{postweakmeasurementstrength}
\end{eqnarray}
then the state in the "no jumping" trajectory will be the same as
initial state. The success probability of retaining the initial
state is given by
\begin{eqnarray}
P^{S}=(1-p)e^{-\Gamma\tau}
 \label{successprobability}
\end{eqnarray}
\section{Generation of two-qubit Bell states when a qubit passes through amplitude damping channel}
Let us start with two qubit entangled state
\begin{eqnarray}
&&|\Psi_{g}^{AA}\rangle=|0\rangle_{1}\otimes(\alpha|0\rangle_{2}+\beta|1\rangle_{2})+|1\rangle_{1}\otimes(\gamma|0\rangle_{2}+\delta|1\rangle_{2}),
\nonumber\\&&|\alpha|^{2}+|\beta|^{2}+|\gamma|^{2}+|\delta|^{2}=1,
\alpha\neq \gamma,\beta\neq \delta
 \label{generaltwoqubit}
\end{eqnarray}
If $\alpha=\gamma$ and $\beta=\delta$ then the state
(\ref{generaltwoqubit}) would become a product state so we have
taken $\alpha\neq \gamma$,$\beta\neq \delta$.\\
Initially both qubit possessed by Alice. Alice then perform a
pre-weak measurement $I\otimes M_{1}$ on the second qubit followed
by feed-forward operation $I\otimes F_{1}$. Alice then send the
second qubit to Bob through amplitude damping channel. The
amplitude damping channel can be described by the Kraus operators
as
\begin{eqnarray}
K_{1}=\begin{pmatrix}
  1 & 0 \\
  0 & \sqrt{1-r}
\end{pmatrix}, K_{2}=\begin{pmatrix}
  0 & \sqrt{r} \\
  0 & 0
\end{pmatrix} \label{adc}
\end{eqnarray}
where $r$ is the magnitude of the decoherence. Taking the argument
from the previous section, the qubit trajectory can be divided
into "jump" and "no jump" trajectory. When the second qubit is in
"jumping" trajectory and Bob operate reversed feed-forward
operation $I\otimes F_{1}^{-1}$ and a partial weak measurement
$I\otimes O_{1}$ on his qubit, the shared state between Alice and
Bob is given by
\begin{eqnarray}
|\Psi_{j}^{AB}\rangle=\frac{1}{\sqrt{2}}(|00\rangle_{12}+|10\rangle_{12})
\label{jumpingtwoqubitAB}
\end{eqnarray}
Clearly the state (\ref{jumpingtwoqubitAB}) is a product state and
it is not useful in any quantum information processing task.\\
When the second qubit is in "no jumping" trajectory and Bob
operate reversed feed-forward operation $I\otimes F_{1}^{-1}$ and
a partial weak measurement $I\otimes O_{1}$ on his qubit, the
shared state between Alice and Bob takes the form
\begin{eqnarray}
|\Psi_{nj}^{AB}\rangle=a|00\rangle_{12}+b|01\rangle_{12}+c|10\rangle_{12}+d|11\rangle_{12},
\label{nojumpingtwoqubitAB}
\end{eqnarray}
where the qubit 1 is with Alice and qubit 2 is with Bob.\\
The parameters $a$, $b$, $c$, $d$ is given by
\begin{eqnarray}
&&a=\frac{\alpha\sqrt{p}\sqrt{1-p_{1}}}{\sqrt{2P^{njN_{1}}}},b=\frac{\beta\sqrt{1-p}e^{\frac{-\Gamma\tau}{2}}}{\sqrt{2P^{njN_{1}}}}\nonumber\\&&
c=\frac{\gamma\sqrt{p}\sqrt{1-p_{1}}}{\sqrt{2P^{nj1N_{1}}}},d=\frac{\delta\sqrt{1-p}e^{\frac{-\Gamma\tau}{2}}}{\sqrt{2P^{nj1N_{1}}}}
\label{parameter}
\end{eqnarray}
where
$P^{nj1N_{1}}=|\gamma|^{2}p(1-p_{1})+|\delta|^{2}(1-p)e^{-\Gamma\tau}$,
$P^{njN_{1}}=|\alpha|^{2}p(1-p_{1})+|\beta|^{2}(1-p)e^{-\Gamma\tau}$.\\
If we assume $1-e^{-\Gamma\tau}=r$ then the decoherence channel is
the same as amplitude damping channel.\\
The Hadamard transformation is given by
\begin{eqnarray}
&&|0\rangle\rightarrow
\frac{1}{\sqrt{2}}(|0\rangle+|1\rangle)\nonumber\\&&
|1\rangle\rightarrow \frac{1}{\sqrt{2}}(|0\rangle-|1\rangle)
\label{hadamard}
\end{eqnarray}
Bob perform Hadamard transformation on his qubit. As a result of
the transformation, the two-qubit state
(\ref{nojumpingtwoqubitAB}) reduces to
\begin{eqnarray}
|\Psi_{njh}^{AB}\rangle&=&\frac{1}{\sqrt{2}}((a-b)|00\rangle_{12}+(a+b)|01\rangle_{12}+(c-d)|10\rangle_{12}\nonumber\\&&+(c+d)|11\rangle_{12}),
\label{nojumpinghadamardtwoqubitAB}
\end{eqnarray}
If we choose $\beta=-\alpha$, $\delta=\gamma$ and the post-weak
measurement strength $p_{1}$ as given in
(\ref{postweakmeasurementstrength}) then the state
$|\Psi_{njh}^{AB}\rangle$ becomes maximally entangled Bell state
i.e. $\frac{1}{\sqrt{2}}(|00\rangle_{12}+|11\rangle_{12})$. The
probability of generating the maximally entangled Bell state with
this procedure is $\frac{1}{2}$.

\section{Generation of three-qubit W-type state when second and third qubit passes through
amplitude damping channel}

In the previous section, we have seen how two-qubit maximally
entangled Bell state is generated when the second qubit is passing
through the amplitude damping channel. In this section, we will
start with this maximally entangled Bell state shared between
Alice and Bob
\begin{eqnarray}
|\psi\rangle_{Bell}^{AB}=\frac{1}{\sqrt{2}}(|00\rangle_{AB}+|11\rangle_{AB})
\label{bellstate}
\end{eqnarray}
Now our task is to generate three-qubit state from two-qubit Bell
state (\ref{bellstate}). To do this, we are using economical
quantum cloning machine. If cloning machine does not need any
ancilla then it is called economical quantum cloning machine.
Economical quantum cloning transformation is given by
\begin{eqnarray}
&&U|0\rangle|0\rangle=|0\rangle|0\rangle\nonumber\\&&
U|1\rangle|0\rangle=cos\alpha|1\rangle|0\rangle+sin\alpha|0\rangle|1\rangle
 \label{eqcm}
\end{eqnarray}
Bob then apply economical quantum cloning transformation to his
qubit. As a result of the cloning transformation, a three-qubit
state is generated and is given by
\begin{eqnarray}
|\psi\rangle_{ABC}&=&\frac{1}{\sqrt{2}}(|000\rangle_{ABC}+cos\alpha|110\rangle_{ABC}\nonumber\\&&+sin\alpha|101\rangle_{ABC})
\label{clonedstate}
\end{eqnarray}
where the qubit $A$ hold by Alice while qubit $B$ and $C$
possessed by Bob.\\
Non-maximally Hadamard transformation is given by
\begin{eqnarray}
&&|0\rangle\rightarrow u|0\rangle+v|1\rangle\nonumber\\&&
|1\rangle\rightarrow v|0\rangle-u|1\rangle,~~~~u^{2}+v^{2}=1
\label{nonmaximallyhadamard}
\end{eqnarray}
Equation (\ref{nonmaximallyhadamard}) reduces to hadamard
transformation when $u=v=\frac{1}{\sqrt{2}}$.\\
Bob apply non-maximally Hadamard transformation on qubit $C$ and
hence the state $|\psi\rangle_{ABC}$ reduces to
\begin{eqnarray}
|\psi\rangle_{ABC}^{nmH}&=&\frac{1}{\sqrt{2}}((|00\rangle_{AB}+cos\alpha|11\rangle_{AB})\otimes
(u|0\rangle_{C}+v|1\rangle_{C})
\nonumber\\&&+sin\alpha|10\rangle_{AB}\otimes(v|0\rangle_{C}-u|1\rangle_{C}))
\label{threequbitstatenmhadamard}
\end{eqnarray}
Bob now executing feed-forward scheme described in section-II to
send the third qubit $C$ to Charlie. When the third qubit $C$
passing through the amplitude damping channel and reached to
Charlie, the three-qubit state shared between Alice, Bob and
Charlie in "no jumping" trajectory becomes
\begin{eqnarray}
|\psi\rangle_{ABC}^{D}&=&u_{1}|000\rangle_{ABC}+u_{2}|001\rangle_{ABC}+u_{3}|110\rangle_{ABC}
\nonumber\\&&+u_{4}|111\rangle_{ABC}+u_{5}|100\rangle_{ABC}\nonumber\\&&+u_{6}|101\rangle_{ABC}
\label{threequbitstatedecoherence}
\end{eqnarray}
where the qubit $C$ possessed by Charlie and the coefficients are
given by
\begin{eqnarray}
&&u_{1}=\frac{u\sqrt{p}\sqrt{1-p_{1}}}{\sqrt{2k_{1}}},u_{2}=\frac{v\sqrt{1-p}e^{\frac{-\Gamma\tau}{2}}}{\sqrt{2k_{1}}},\nonumber\\&&
u_{3}=\frac{ucos\alpha\sqrt{p}\sqrt{1-p_{1}}}{\sqrt{2k_{1}}},
u_{4}=\frac{vcos\alpha\sqrt{1-p}e^{\frac{-\Gamma\tau}{2}}}{\sqrt{2k_{1}}},\nonumber\\&&
u_{5}=\frac{vsin\alpha\sqrt{p}\sqrt{1-p_{1}}}{\sqrt{2k_{2}}},
u_{6}=\frac{-usin\alpha\sqrt{1-p}e^{\frac{-\Gamma\tau}{2}}}{\sqrt{2k_{2}}}\nonumber\\&&
k_{1}=u^{2}p(1-p_{1})+v^{2}(1-p)e^{-\Gamma\tau},\nonumber\\&&
k_{2}=v^{2}p(1-p_{1})+u^{2}(1-p)e^{-\Gamma\tau}
\label{threequbitstateparameter}
\end{eqnarray}
3-tangle of the state $|\psi\rangle_{ABC}^{D}$ is zero and hence
it represent a $W-type$ state which is shared between Alice, Bob and Charlie.\\
Charlie apply Hadamard transformation on his qubit and the
resultant three-qubit state takes the form as
\begin{eqnarray}
|\psi\rangle_{ABC}^{DH}&=&\frac{1}{\sqrt{2}}[(u_{1}+u_{2})|000\rangle_{ABC}+(u_{1}-u_{2})|001\rangle_{ABC}\nonumber\\&&+(u_{3}+u_{4})|110\rangle_{ABC}
+(u_{3}-u_{4})|111\rangle_{ABC}\nonumber\\&&+(u_{5}+u_{6})|100\rangle_{ABC}\nonumber\\&&+(u_{5}-u_{6})|101\rangle_{ABC}]
\label{threequbitstatedecoherence}
\end{eqnarray}
If we choose the post-weak measurement strength as
\begin{eqnarray}
p_{1}=1-\frac{v^{2}(1-p)e^{-\Gamma\tau}}{u^{2}p} \label{pwms1}
\end{eqnarray}
then $u_{1}=u_{2}=\frac{1}{2}$, $u_{3}=u_{4}=\frac{cos\alpha}{2}$,
$u_{5}=\frac{v^{2}sin\alpha}{\sqrt{2(u^{4}+v^{4})}}$,
$u_{6}=\frac{-u^{2}sin\alpha}{\sqrt{2(u^{4}+v^{4})}}$. This
reduces the state $|\psi\rangle_{ABC}^{DH}$ to
\begin{eqnarray}
|\psi\rangle_{ABC}^{W}&=&\frac{1}{\sqrt{N}}[\frac{1}{\sqrt{2}}|000\rangle_{ABC}+\frac{1}{\sqrt{2}}cos\alpha|110\rangle_{ABC}
\nonumber\\&&+\frac{(v^{2}-u^{2})sin\alpha}{2\sqrt{u^{4}+v^{4}}}|100\rangle_{ABC}\nonumber\\&&+
\frac{sin\alpha}{2\sqrt{(u^{4}+v^{4})}}|101\rangle_{ABC}]
\label{threequbitwstate}
\end{eqnarray}
where
$N=\frac{1}{2}+\frac{cos^{2}\alpha}{2}+\frac{((v^{2}-u^{2})^{2}+1)sin^{2}\alpha}{4(v^{4}+u^{4})}$.\\
For $u=v=\frac{1}{\sqrt{2}}$, the three-qubit state
(\ref{threequbitwstate}) reduces to
\begin{eqnarray}
|\psi\rangle_{ABC}^{W1}&=&\frac{1}{\sqrt{2}}[|000\rangle_{ABC}+cos\alpha|110\rangle_{ABC}
\nonumber\\&&+sin\alpha|101\rangle_{ABC}]
\label{threequbitwstate1}
\end{eqnarray}
When Alice apply the Pauli operator $\sigma_{x}$ on her qubit then
the state $|\psi\rangle_{ABC}^{W1}$ reduces to the state
introduced by Agrawal and Pati \cite{ap}
\begin{eqnarray}
|\psi\rangle_{ABC}^{W2}&=& (\sigma_{x}\otimes I\otimes I
)|\psi\rangle_{ABC}^{W1}\nonumber\\&&=
\frac{1}{\sqrt{2}}[|100\rangle_{ABC}+cos\alpha|010\rangle_{ABC}
\nonumber\\&&+sin\alpha|001\rangle_{ABC}]
\label{threequbitwstate2}
\end{eqnarray}
This class of $W-type$ states can be used for perfect
teleportation and superdense coding.

\section{Perfect teleportation of two non-orthogonal states with $|\psi\rangle_{ABC}^{W2}$ }
In this section we have shown that how two non-orthogonal states
can be teleported via a three-qubit state
$|\psi\rangle_{AAB}^{W2}$.\\
Let us consider two nonorthogonal states to be teleported is given
by
\begin{eqnarray}
|\chi_{1}\rangle_{A}=x|0\rangle_{A}+y|1\rangle_{A},~~x^{2}+y^{2}=1
\label{non-orthogonalstate1}
\end{eqnarray}
\begin{eqnarray}
|\chi_{2}\rangle_{A}&=&(sx+y\sqrt{1-s^{2}})|0\rangle_{A}\nonumber\\&&+(sy-x\sqrt{1-s^{2}})|1\rangle_{A}
\label{non-orthogonalstate2}
\end{eqnarray}
We note that $\langle\chi_{1}|\chi_{2}\rangle=s,~~0\leq s \leq 1
$. Let us assume that the two single qubit non-orthogonal states
defined above are with Alice and she has information about the
parameters $x$ and $s$. She want to teleport the messages encoded
in the non-orthogonal states to Bob with the help of three-qubit
$W-type$ state
\begin{eqnarray}
|\psi\rangle_{AAB}^{W2}&=&
=\frac{1}{\sqrt{2}}[|100\rangle_{AAB}+cos\alpha|010\rangle_{AAB}
\nonumber\\&&+sin\alpha|001\rangle_{AAB}]
\label{threequbitwstate1}
\end{eqnarray}
where the first two qubits are with Alice and the third qubit is
with Bob. $\alpha$ is the economical cloning machine parameter.\\
The composite five qubit state is given by
\begin{eqnarray}
|\psi\rangle_{AAAAB}=|\chi_{1}\rangle_{A}\otimes|\chi_{2}\rangle_{A}\otimes
|\psi\rangle_{AAB}^{W2} \label{fivequbitstate}
\end{eqnarray}
Alice then perform two Bell state measurements on her qubits. As a
result of the measurement, either a bit or a qubit is generated at
Bob's site. Appearance of bit at Bob's site means no
non-orthogonal states appeared at his place and hence this case is
considered as failure of the protocol. But the case, when qubit is
generated, can be considered as success of the protocol because in
this case the generated qubit can be converted into one of the two
non-orthogonal states $|\chi_{1}\rangle_{B}$ or $|\chi_{2}\rangle_{B}$.\\
Let us consider the following cases:\\
\textbf{Case-Ia:} If the measurement outcome is
$|\psi^{+}\rangle_{AA}\otimes |\phi^{+}\rangle_{AA}$ or
$|\psi^{+}\rangle_{AA}\otimes |\phi^{-}\rangle_{AA}$ then Alice
sent two classical bits $|0\rangle\otimes|0\rangle$ to Bob. Bob
applies $I$ on the received qubit after getting two classical bits
from Alice. Before Bell state measurement, if Alice chooses the
cloning machine parameter $\alpha$ in such a way that
$sin\alpha=\frac{K^{2}-1}{K^{2}+1}$ and
$cos\alpha=\frac{-2K}{K^{2}+1}$, where
$K=\frac{x(sx+y\sqrt{1-s^{2}})}{y(sy-x\sqrt{1-s^{2}})}$ then the
qubit appear at Bob's place is $|\chi_{1}\rangle_{B}$.\\
\textbf{Case-Ib:} If the measurement outcome is
$|\psi^{+}\rangle_{AA}\otimes |\phi^{+}\rangle_{AA}$ or
$|\psi^{+}\rangle_{AA}\otimes |\phi^{-}\rangle_{AA}$ then Alice
sent two classical bits $|0\rangle\otimes|1\rangle$ to Bob. After
getting classical bits, Bob applies $\sigma_{x}$ on the received
qubit. If Alice chooses the cloning machine parameter $\alpha$ in
such a way that $sin\alpha=\frac{\sqrt{2-K^{2}}+K}{2}$ and
$cos\alpha=\frac{\sqrt{2-K^{2}}-K}{2}$ and  then the
qubit appear at Bob's place is $|\chi_{2}\rangle_{B}$.\\
\textbf{Case-IIa:} If the measurement outcome is
$|\psi^{-}\rangle_{AA}\otimes |\phi^{+}\rangle_{AA}$ or
$|\psi^{-}\rangle_{AA}\otimes |\phi^{-}\rangle_{AA}$ then Alice
sent two classical bits $|1\rangle\otimes|0\rangle$ to Bob. Bob
then applies $\sigma_{z}$ on the received qubit. If Alice chooses
$sin\alpha=\frac{K^{2}-1}{K^{2}+1}$ and
$cos\alpha=\frac{2K}{K^{2}+1}$ before Bell state measurement then
in this case also the state appears at Bob's site is $|\chi_{1}\rangle_{B}$.\\
\textbf{Case-IIb:} If the measurement outcome is
$|\psi^{-}\rangle_{AA}\otimes |\phi^{+}\rangle_{AA}$ or
$|\psi^{-}\rangle_{AA}\otimes |\phi^{-}\rangle_{AA}$ then Alice
sent a classical bit $|1\rangle\otimes|1\rangle$ to Bob. After
getting classical bits, Bob applies $-i\sigma_{y}$ on the received
qubit. If Alice chooses $sin\alpha=\frac{K+\sqrt{2-K^{2}}}{2}$ and
$cos\alpha=\frac{K-\sqrt{2-K^{2}}}{2}$ before Bell state
measurement then the state appears at Bob's site is $|\chi_{2}\rangle_{B}$.\\
\textbf{Case-IIIa:} If the measurement outcome is
$|\psi^{+}\rangle_{AA}\otimes |\psi^{+}\rangle_{AA}$ or
$|\psi^{+}\rangle_{AA}\otimes |\psi^{-}\rangle_{AA}$ then Alice
sent two classical bits $|0\rangle\otimes|0\rangle$ to Bob. Bob
applies $I$ on the received qubit After getting two classical bits
from Alice. If Alice Chooses
$sin\alpha=\frac{\sqrt{2-L^{2}}+L}{2}$ and
$cos\alpha=\frac{\sqrt{2-L^{2}}-L}{2}$, where
$L=\frac{x(sy-x\sqrt{1-s^{2}})}{y(sx+y\sqrt{1-s^{2}})}$ before
Bell state measurement then the
state appears at Bob's site is $|\chi_{2}\rangle_{B}$.\\
\textbf{Case-IIIb:} If the measurement outcome is
$|\psi^{+}\rangle_{AA}\otimes |\psi^{+}\rangle_{AA}$ or
$|\psi^{+}\rangle_{AA}\otimes |\psi^{-}\rangle_{AA}$ then Alice
sent two classical bits $|0\rangle\otimes|0\rangle$ to Bob. Bob
applies $I$ on the received qubit After getting two classical bits
from Alice. If Alice Chooses in this case
$sin\alpha=\frac{L^{2}-1}{L^{2}+1}$ and
$cos\alpha=\frac{-2L}{L^{2}+1}$ before Bell state measurement then
the state appears at Bob's site is $|\chi_{1}\rangle_{B}$.\\
\textbf{Case-IVa:} If the measurement outcome is
$|\psi^{-}\rangle_{AA}\otimes |\psi^{+}\rangle_{AA}$ or
$|\psi^{-}\rangle_{AA}\otimes |\psi^{-}\rangle_{AA}$ then Alice
sent two classical bits $|1\rangle\otimes|0\rangle$ to Bob. Bob
applies $\sigma_{z}$ on the received qubit after getting two
classical bits from Alice. If Alice Chooses
$sin\alpha=\frac{L+\sqrt{2-L^{2}}}{2}$ and
$cos\alpha=\frac{L-\sqrt{2-L^{2}}}{2}$ before Bell state
measurement then the state appears at Bob's site is $|\chi_{2}\rangle_{B}$.\\
\textbf{Case-IVb:} If the measurement outcome is
$|\psi^{-}\rangle_{AA}\otimes |\psi^{+}\rangle_{AA}$ or
$|\psi^{-}\rangle_{AA}\otimes |\psi^{-}\rangle_{AA}$ then Alice
sent two classical bits $|1\rangle\otimes|0\rangle$ to Bob. Bob
applies $\sigma_{z}$ on the received qubit after getting two
classical bits from Alice. If Alice Chooses
$sin\alpha=\frac{L^{2}-1}{L^{2}+1}$ and
$cos\alpha=\frac{2L}{L^{2}+1}$ before Bell state measurement then
the state appears at Bob's site is $|\chi_{1}\rangle_{B}$.
\section{Conclusion}
To summarize, we have made a study of the generation of a pure
two-qubit Bell states $\frac{1}{\sqrt{2}}(|00\rangle+|11\rangle)$
shared between two distant partners, even though a qubit from a
two-qubit general state is interacting with the noisy environment.
This would be possible only if we use C-Q Wang et.al.
weak-measurement-based feed-forward control scheme. A three-qubit
W-type states $\frac{1}{\sqrt{2}}[|100\rangle+cos\alpha|010\rangle
+sin\alpha|001\rangle]$, where $\alpha$ is the economical quantum
cloning machine parameter, can also be prepared by economical
quantum cloning machine and feed-forward scheme. We have shown
that the generated three-qubit state shared between Alice, Bob and
Charlie reside in three different laboratories. Since the two
-qubit Bell states and a particular form of three-qubit W-type
states discussed in this work are very useful in secret sharing
and quantum cryptography so the study of their production is very
important in quantum information theory. Further we have shown
that one of the two non-orthogonal states can be teleported via
W-type states.

\end{document}